\documentstyle[12pt,epsf]{article}

\newcommand{\abstracts}[1]{{
\centering{\begin{minipage}{12.2truecm}
\normalsize\baselineskip=15pt
\centerline{\footnotesize ABSTRACT}\vspace*{0.3cm}
\parindent=20pt #1
\end{minipage}}\par}}

\newcommand{\beq}{\begin{equation}}
\newcommand{\eeq}{\end{equation}}
\newcommand{\beqn}{\begin{eqnarray}}
\newcommand{\eeqn}{\end{eqnarray}}
\newcommand{\summ}[2]{\sum\limits_{#1}^{#2}}

\newcommand{\Z}{{Z \!\!\! Z}}

\newcommand{\dual}{\mbox{}^{\ast}}
\newcommand{\diff}{\partial}
\newcommand{\eq}[1]{(\ref{#1})}
\def\NP{ Nucl.~Phys.}
\def\PL{ Phys.~Lett.}
\def\PR{ Phys.~Rev.}

\begin{document}

\thispagestyle{empty}

\begin{center}
\vspace{-1cm}
\begin{flushright}
{\large FISIST/17-98/CFIF}\\
\vskip 2mm
{\large ITEP-TH-70/97}
\end{flushright}
\vspace{1.5cm}
{\baselineskip=24pt
{\LARGE \bf Abelian Dyons \\
in the Maximal Abelian Projection \\
of $SU(2)$ Gluodynamics}\\

{\baselineskip=24pt

\vspace{1cm}

{\large M.N.~Chernodub, F.V.~Gubarev and M.I.~Polikarpov}\\

\vspace{.1cm}
{ \it
ITEP, B.Cheremushkinskaya 25, Moscow, 117259, Russia

}
}
}
\end{center}

\vspace{.5cm}

\abstracts{
Correlations of the topological charge~$Q$, the
electric current~$J^e$ and the magnetic current~$J^m$ in $SU(2)$
lattice gauge theory in the Maximal Abelian projection are
investigated.  It occurs that the correlator $\ll QJ^eJ^m \gg$ is
nonzero for a wide range of values of the bare charge.  It is
shown that:  {\it (i)} the abelian monopoles in the Maximal
Abelian projection are dyons which carry {\it fluctuating}
electric charge; {\it (ii)} the sign of the electric charge
$e(x)$ coincides with that of the product of the monopole charge
$m(x)$ and the topological charge density $Q(x)$.}

\vspace{0.3cm}


There are several approaches to the confinement problem in
QCD~\cite{Simonov}.  The most popular is the model of
the dual superconducting vacuum~\cite{MatH76}:  the vacuum is
supposed to be a media of condensed magnetic charges (monopoles).
This model naturally explains the confinement phenomena.  During
the last decade the method of abelian projections~\cite{tH81} has been
successfully used in lattice calculations in order to show that the
vacuum of gluodynamics behaves as a dual
superconductor (see {\it e.g.} the reviews~\cite{LatticeRevievs}).
Most of the  numerical simulations were  performed in the so called
Maximal Abelian (MaA) projection~\cite{KrScWi87}.

An oldest and rather popular model of the QCD vacuum is the
instanton--anti-instan\-ton medium (see~\cite{InstantonicReviews}
and the references therein).  It is not clear whether the
confinement phenomenon can be explained within this
approach~\cite{InstKucha}.  However, the instanton--based models
may have some relation to the dual superconductor model, since
the instantons and monopoles are interrelated, as demonstrated
analytically in~\cite{Inst1} and numerically
in~\cite{Inst2,BornSchierholz}.  One can expect that the
instantons may affect the properties of the abelian monopoles in
the MaA projection.  Indeed, it has been shown by numerical
calculations~\cite{BornSchierholz} that the abelian monopole
becomes the abelian dyon in the field of a single instanton.

In this paper we study the electric charge of the abelian
monopoles in the vacuum of lattice $SU(2)$ gluodynamics.
The numerical data presented below show that the
abelian monopole in the maximal abelian projection carry
fluctuating electric charge.  This fact is very important for the
dynamics of the confining strings.  It was shown in the recent
paper \cite{ABdyon}, that in the theory with condensed dyons the
topological interaction exists for the expectation value of the
Wilson loop.  This long range interaction leads to the nontrivial
dynamics of the confining string spanned on the Wilson loop and
may induce a long range potential between quark and antiquark.


We start the discussion of the electric charge of the monopole
with the simple example. Consider the monopole currents on the
(anti)instanton background \cite{BornSchierholz}.
The (anti-) self-dual gauge fields satisfy the equation:
\beqn
F_{\mu\nu}(A)=\pm \frac{1}{2} \varepsilon_{\mu\nu\alpha\beta}
F_{\alpha\beta}(A) \equiv \pm \dual F_{\mu\nu}\,,
\label{duality}
\eeqn
where $F_{\mu\nu}(A) =
\partial_{[\mu,} A_{\nu]} + i [A_\mu,A_\nu]$.
The numerical calculations performed in Ref.\cite{BornSchierholz}
show that the
 abelian monopole current is accompanied by the electric
current.

The correlation of the electric and the
magnetic currents in the field of
the instanton can be qualitatively explained as
follows~\cite{BornSchierholz}. The MaA projection is
defined~\cite{KrScWi87} by the minimization of the functional
$R[A^{\Omega}(x)]$ over  the gauge transformations $\Omega(x)$, where
$R[A]=\int d^4 x [(A_{\mu}^{1})^2 + (A_{\mu}^{2})^2]$. Therefore, in
this projection the off--diagonal gauge fields $A^\pm_\mu = A^1_\mu
\pm i A^2_\mu$ are suppressed with respect to the diagonal gauge field
$A^3_\mu$. Thus,  the commutator term   $1/2 Tr(\sigma^3
[A_{\mu},A_{\nu}]) =\varepsilon^{3bc}A^b_{\mu}A^c_{\nu}$
in $F_{\mu\nu}^3$
 is small compared
to the abelian field-strength $f_{\mu\nu}(A)=\diff_{[\mu,} A^3_{\nu ]}$.
Therefore, in the MaA projection eq.(\ref{duality}) yields:
\beqn
f_{\mu\nu}(A)\approx\pm\dual f_{\mu\nu}(A)\,.  \label{abelian_duality}
\eeqn
Thus the abelian monopole currents must be
correlated with the electric currents:
\beqn J^e_{\mu}=\diff_{\nu}
f_{\mu\nu}(A)\approx \pm \diff_{\nu}\dual f_{\mu\nu}(A)=J^m_{\mu}\,.
\label{jm_je}
\eeqn
Therefore, in the MaA projection the abelian monopoles
are dyons in the background of (anti) self-dual $SU(2)$ fields.

In the present publication we study the correlation of electric and
magnetic currents in the vacuum  of $SU(2)$ lattice gluodynamics
in the MaA projection. The definition of the abelian monopole current
on the lattice is~\cite{DGT}:
\beqn
J^m_{\mu}(y)=\frac{1}{4\pi} \sum\limits_{\nu\lambda\rho}
\varepsilon_{\mu\nu\lambda\rho}
[\bar\theta_{\lambda\rho}(x+\hat{\mu})-\bar\theta_{\lambda\rho}(x)].
\label{monopole_current}
\eeqn
Here the function $\bar\theta_{\mu\nu}$ is the normalized plaquette
angle $\theta_{\mu\nu}$: $\bar\theta_{\mu\nu}=
\theta_{\mu\nu} - 2\pi k_{\mu\nu}\in (-\pi;\pi]$, $k_{\mu\nu} \in
\Z$. The monopole current $J^m_\mu (x)$ is attached to the links of
the dual lattice.  One can easily show that the monopole currents
are quantized, $J^m_{\mu}\in \Z$, and conserved, $\diff_{\mu}
J^m_{\mu}=0$.

The lattice electric current is defined as \cite{BornSchierholz}:
\beqn
K^e_{\mu}(x)=\frac{1}{2\pi} \sum\limits_{\nu}
[\bar\theta_{\mu\nu}(x)-\bar\theta_{\mu\nu}(x-\hat{\nu})].
\label{electric_current}
\eeqn
In the continuum limit,  this equation corresponds to the usual
definition: $K^e_\mu=\diff_{\nu} f_{\mu\nu}$. The electric currents
$K^e_{\mu}$ are defined on the original lattice.  They are conserved,
i.e., $\diff_{\mu} K^e_{\mu}=0$, but, contrary to the magnetic currents,
are not quantized.

In order to calculate the correlators of the electric and
the magnetic
currents,  one has to define the electric current on the dual lattice
or the magnetic current on the original lattice.  We use  the following
definition of the electric current $J^e_\mu$ on the {\it dual}
lattice:
\beqn
J^e_{\mu}(y) = \frac{1}{16}\summ{x\in \dual C(y,\mu)}{}
\left[ K^e_{\mu}(x)+K^e_{\mu}(x-\hat{\mu})\right]\,,
\label{correspondence}
\eeqn
where the summation in the {\it r.h.s.} is over the
 eight vertices $x$ of the
3-dimensional cube $\dual C(y,\mu)$, to which the current
$J^e_{\mu}(y)$ is dual. As in eq.\eq{monopole_current}, the point $y$
lies on the dual lattice and the vertices $x$ belong to the original
lattice. The current $J^e_\mu$ defined by eq.\eq{correspondence} has
the standard continuum limit:  $J^e_{\mu}=\diff_{\nu} f_{\mu\nu}$.

For the topological charge density we use the simplest definition:
\beqn
 Q(x)=\frac{1}{2^9 \pi^2}
\sum_{\mu_1,\mu_2,\mu_3,\mu_4=-4}^{4}
\varepsilon^{\mu_1,\mu_2,\mu_3,\mu_4}
Tr[U_{\mu_1\mu_2}(x) U_{\mu_3\mu_4}(x)]\,,
\eeqn
where $U_{\mu_1\mu_2}(x)$ is the plaquette matrix. On the dual
lattice the topological charge density $Q(y)$ corresponding to the
monopole current $J^m_{\mu}(y)$ is defined by averaging over
the sites nearest to the current $J^m_{\mu}(y)$:
\beqn
Q(y)=\frac{1}{8}\summ{x}{} Q(x)\;;
\eeqn
the summation here is the same as in~eq.(\ref{correspondence}).

The simplest (connected) correlator of the electric and
the magnetic
currents is: $\ll J^m_{\mu}J^e_{\mu} \gg =
<J^m_{\mu}J^e_{\mu}> - < J^m_{\mu}> <J^e_{\mu}> \equiv
<J^m_{\mu}J^e_{\mu}>\,,$ $<J^m_{\mu}>=<J^e_{\mu}>=0$ is due to the Lorentz
invariance.  The correlator $<J^m_{\mu}J^e_{\mu}>$ is zero due to the
opposite parities of the operators $J^m$ and $J^e$.

The nonvanishing correlator is $\ll J^m_{\mu}(y) J^e_{\mu}(y) Q(y)
\gg$ which is both Lorentz and parity invariant. Due to equalities
$<J^m_{\mu}(y)> = <J^e_{\mu}(y)> = <Q(y)> = 0$ the connected
correlator is:
\beqn
G = \ll J^m_{\mu}(y) J^e_{\mu}(y) Q(y) \gg =
<J^m_{\mu}(y) J^e_{\mu}(y) Q(y)>\,.
\label{CorrelatorOne}
\eeqn
The density of electric and magnetic charges strongly depends on
$\beta$. To compensate this dependence we consider the normalized
correlator $\bar G$:
\beqn
{\bar G } =
\frac{1}{\rho^e \rho^m} \, <J^m_{\mu}(y) J^e_{\mu}(y) q(y)>\,,
\label{G}
\eeqn
where
\beqn
\rho_{m,e} = \frac{1}{4V} \summ{l}{} <|J^{m,e}_l|>\,, \qquad
q (x) = \frac{Q(y)}{|Q(y)|} \equiv {\rm sign} Q(x) \,, \nonumber
\eeqn
  $V$ being the lattice volume (total number of sites).


We perform the numerical simulations on the $8^4$ lattice with
periodic boundary conditions. We thermalize lattice fields using the
standard heat bath algorithm. All correlators for each value of
$\beta$ were calculated on 100 statistically independent
configurations. To fix the MaA projection we use the
overrelaxation algorithm of Ref.~\cite{MaOg90}.

\begin{figure}[htb]
\vspace{0.7cm}
\centerline{\epsfxsize=0.6\textwidth\epsfbox{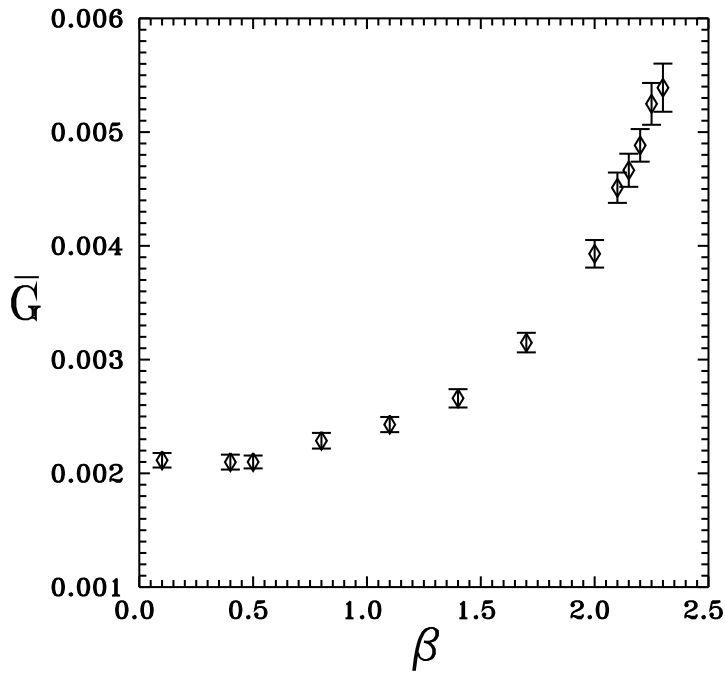}}
\caption{The correlator ${\bar G}$, eq.\eq{G}, $vs.$ $\beta$.}
\label{one}
\end{figure}

The correlator ${\bar G}$ given by eq.\eq{G}  $vs.$ $\beta$ is shown in
Figure~\ref{one}. Since the product of electric and
magnetic currents is correlated with the topological charge,
we see that   the abelian monopole carries the electric charge which
depends on the topological charge density at the abelian monopole
position.

We have found that the correlator ${\bar G}$ grows during the cooling of the
field configurations. This means that the strongest correlation of the
electric and the
magnetic charges is observed in  (anti-) self-dual fields
(e.g.,  for the instanton configuration).

\begin{figure}[htb]
\centerline{\epsfxsize=0.6\textwidth\epsfbox{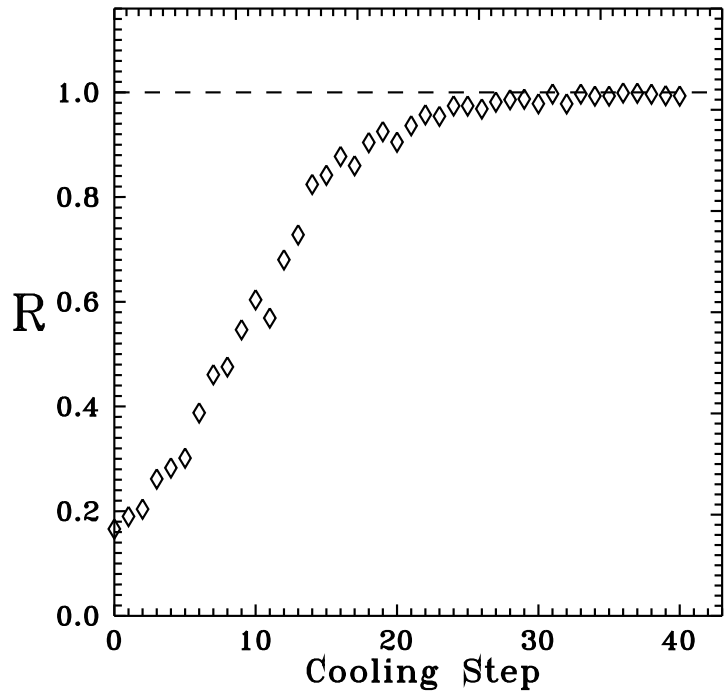}}
\caption{The ratio $R$, eq.\eq{R}, $vs.$ the number of cooling steps
at $\beta=2.2$.}
\label{three}
\end{figure}

In order clarify how our results are related to  those of
Ref.~\cite{BornSchierholz}, we study the correlator
\beqn
R = \frac{<J^m_{\mu}(y) J^e_{\mu}(y) q(y)>}{<
|J^m_{\mu}(y) J^e_{\mu}(y) q(y)|>}\,
\label{R}
\eeqn
in the cooled vacuum. The correlator $R$ $vs.$ the number of the
cooling steps  $n$  is shown in Figure~\ref{three} at $\beta=2.2$. The
plateau $R=1$ at $n > 25$ corresponds to the classical instanton
configuration   studied in Ref.~\cite{BornSchierholz}. In the
real (not cooled) vacuum,  the field configurations are not
self--dual, and we have  $R<1$ at $n=0$.


Thus our results show that the abelian monopoles in the MaA projection of
$SU(2)$ gluodynamics carry a  fluctuating electric charge. The sign of
the electric charge is equal to that  of the product of the
topological charge density and the magnetic charge. The large
electric charge is in the (anti-) self--dual vacuum, while in the
real (not cooled) vacuum the induced charge is smaller.

M.I.P. acknowledges the kind hospitality of the Theoretical
Department of the Kanazawa University and of Centro
de F\'\i sica das Interac\c c\~oes Fundamentais, Edif\'\i cio
Ci\^encia, Instituto Superior T\'ecnico (Lisboa) where a part of
this work has been done.  F.V.G.  is grateful for the kind
hospitality of the Theoretical Physics Department of the Free
University of Amsterdam. This work was partially
supported by the grants INTAS-RFBR-95-0681, INTAS-96-370,
RFBR-96-15-96740 and RFBR-96-02-17230a; M.N.Ch. was supported by the
INTAS Grant 96-0457 within the research program of the International
Center for Fundamental Physics in Moscow.

\end{document}